\documentclass[11pt]{article}
\usepackage{authblk}
\usepackage{fullpage}
\usepackage{graphicx}
\usepackage{amsmath}

\usepackage{breakcites}
\usepackage[round]{natbib} 

\usepackage{amssymb}
\usepackage{amsthm}
\usepackage{tikz}

\usepackage{enumitem}
\usepackage[utf8]{inputenc} % allow utf-8 input
\usepackage[T1]{fontenc}    % use 8-bit T1 fonts
\usepackage{hyperref}       % hyperlinks
\usepackage{url}            % simple URL typesetting
\usepackage{svg}
\usepackage{booktabs}       % professional-quality tables
\usepackage{amsfonts}       % blackboard math symbols
\usepackage{microtype}      % microtypography
\usepackage{xcolor}         % colors
\usepackage[export]{adjustbox}
\usepackage{dblfloatfix}
\usepackage[mode=buildnew]{standalone}% requires -shell-escape
\usepackage{caption}
\usepackage{subcaption}
\usepackage[bottom]{footmisc}
\usepackage{float}
\usepackage{wrapfig}
\usepackage{titletoc}

\renewcommand{\cite}{\citep}

\hypersetup{
    colorlinks=true,
    linkcolor=black,
    citecolor=cyan,
    urlcolor=black,
}

\title{A broken duet: \\
multistable dynamics of dyadic interactions}
\author[1,*]{Johan Medrano}
\author[2,*]{Noor Sajid}
\affil[1]{Wellcome Centre for Human Neuroimaging, University College London, UK}
\affil[2]{Max Planck Institute for Biological Cybernetics, Tübingen, Germany}
\affil[*]{These authors contributed equally to this work.}

\date{}

\begin{document}

\maketitle

\begin{abstract}
    Misunderstandings in dyadic interactions often persist despite our best efforts, particularly between native and non-native speakers, resembling a broken duet that refuses to harmonise. This paper delves into the computational mechanisms underpinning these misunderstandings through the lens of the broken Lorenz system—a continuous dynamical model. By manipulating a specific parameter regime, we induce bistability within the Lorenz equations, thereby confining trajectories to distinct attractors based on initial conditions. This mirrors the persistence of divergent interpretations that often result in misunderstandings. Our simulations reveal that differing prior beliefs between interlocutors result in misaligned generative models, leading to stable yet divergent states of understanding when exposed to the same percept. Specifically, native speakers equipped with precise (i.e., overconfident) priors expect inputs to align closely with their internal models, thus struggling with unexpected variations. Conversely, non-native speakers with imprecise (i.e., less confident) priors exhibit a greater capacity to adjust and accommodate unforeseen inputs. Our results underscore the important role of generative models in facilitating mutual understanding (i.e., establishing a shared narrative) and highlight the necessity of accounting for multistable dynamics in dyadic interactions.
\end{abstract}

\setcounter{section}{0}
\section{Introduction}\label{sec:intro}

Understanding how individuals interpret the intentions of others is a fundamental challenge in studying dyadic interactions, particularly within the complex dynamics of interactions marred by linguistic differences, such as syntax and pronunciation~\citep{gopnik1990linguistic,dkabrowska2006individual}. These challenges are often exacerbated in interactions between native and non-native speakers, who may have significantly different prior experiences with the language in question~\citep {long1983linguistic,dkabrowska2006individual}. Consequently, misunderstandings between native and non-native speakers can impede effective dialogue~\citep{clahsen2006native}. To investigate this phenomenon, we adopt a Bayesian approach, positing that misunderstandings between listeners of different proficiency levels result from asynchronous predictions about causes of outcomes under disparate model structures (or priors)~\citep{friston2009free,penny2012}.

\begin{figure}[t!]
 \centering
    \includegraphics[width=1\textwidth]{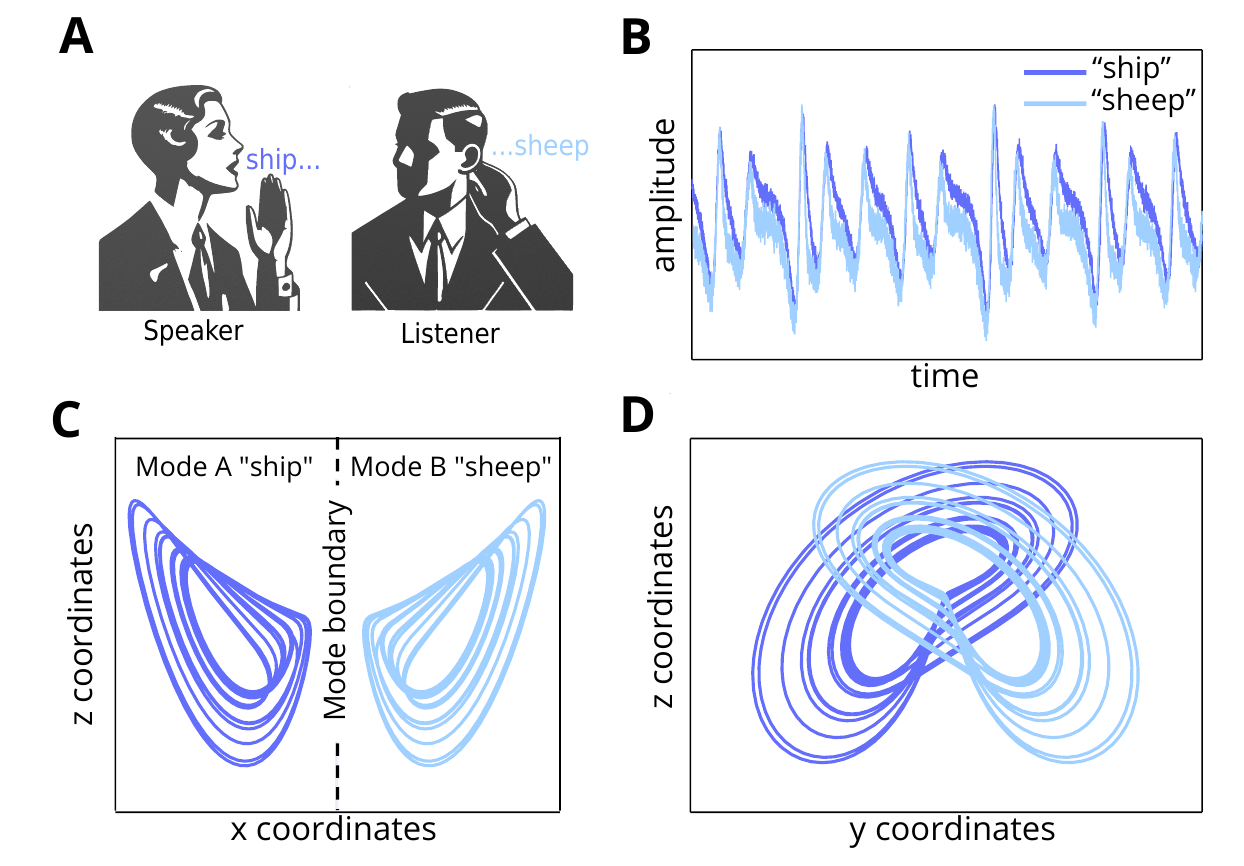} 
    \caption{Visual illustration of the problem setting. \textbf{A} Two agents (speaker and listener) are communicating and there is a misunderstanding between them regarding the word "sheep". \textbf{B} We present the percept measured in amplitude (y-axis) across time (x-axis) for both the pronounced word (by the speaker) and the inferred word (by the listener). \textbf{C-D} The plots illustrate the internal beliefs across the ($x$, $y$, and $z$) coordinate space for a broken Lorenz system. In \textbf{C}, across the $x$ (x-axis) and $z$ (y-axis) coordinate space, the two words "sheep" and "ship" are separated along the $x$ coordinates. Conversely, in \textbf{D}, across the $y$ (x-axis) and $z$ (y-axis) coordinate space, the separation is less trivial. Accordingly, modelling the percept as a weighted sum of these coordinates complicates mode recognition from the percept, leading to misunderstanding of the heard word.}
    \label{fig:summary}
\end{figure}

% Shared narrative - 
For this, we assume that individuals model the causes of their sensory data and refine these models to maximise Bayesian model evidence~\citep{Schmidhuber:92ncchunker,dayan1995helmholtz,knill2004}. Accordingly, effective exchanges manifest through generalised synchrony when individuals adopt the same generative model of communicative behaviour, thereby establishing a shared narrative~\citep{friston2015duet,friston2015active,friston2020generative}. This concept is grounded within the framework of predictive coding and active inference~\citep{friston2020generative,sajid2022active,friston2023free}, where individuals are not necessarily predicting each other's specific actions but rather predict their responses to the incoming percept under a mutually shared narrative~\citep{clark2013whatever,friston2020generative}. Thus, synchronisation fails when there are divergences in the generative models employed by the interacting individuals~\citep{schwartenbeck2015optimal}. Under such circumstances, the prediction mechanisms become less effective, leading to a breakdown in communication~\citep{friston2021active,friston2020generative}. This breakdown can be viewed not merely as a failure of linguistic exchange but as a fundamental misalignment of underlying model structures. 

% Our main point - effective communication:
%  Understanding can be seen as getting in the right attracting state i.e., getting the right meaning
To understand how communication is hindered in the absence of an underlying shared narrative -- formalised by the generative process and model --  we consider dyadic interactions modelling using multistable systems~\citep{lorenz1963deterministic,li2014multistability}. These multistable systems are characterised by several disjoint attracting sets with different basin of attractions. Each attracting set can be seen as capturing a stable mode, i.e., an interpretation of or belief about the sensory data~\citep{tognoli2014metastable, mcintosh2019hidden,medrano2024linking}. Under this perspective, multistable systems carry distinct interpretations (or beliefs) that can coexist based on some incoming sensory information (Fig~\ref{fig:summary}). Importantly, this serves as a way to illustrate that computational mechanisms of understanding are reliant not only on appropriate perceptual inference but also on convergence towards the appropriate attractor set. This implies that the efficacy of linguistic interactions between native and non-native speakers depends on ($i$) the ability to infer the percept (i.e., the auditory waveform being sampled from the environment), and ($ii$) the capacity to assign semantic meaning to the inferred percept.

% What we show:
To elucidate our approach, we consider the communication dynamics between native and non-native speakers across two scenarios. The first scenario contends with the interaction between a native listener and a non-native speaker. For this, we introduce a 'native listener' model with incredibly precise (or overly confident) priors about the incoming sensory data. This model, characterised by high predictive confidence, expects inputs to conform to its beliefs (e.g., syntax, pronunciation). These lead to efficient communication in familiar contexts but introduce challenges in unexpected variations, like unconventional syntax or pronunciation errors. Thus, the native listener might experience internal confusion when hearing a non-native speaker say, "Yesterday I go to the store", due to the rigid expectation of "went" instead of "go" despite perceptual synchronisation. For the second scenario, we consider the interaction between two non-native individuals. The non-native listener is modelled with flexible (i.e., less confident) priors that can be adjusted to accommodate unexpected inputs more effectively. Thus, the non-native listener would not experience the same internal confusion when hearing another speaker say, "Yesterday I go to the store". The two scenarios highlight that establishing a shared narrative and avoiding linguistic misunderstandings—beyond mere perceptual inference—necessitates an appropriate amount of flexibility to enable the synchronisation of internal beliefs.

%% takeaway: 
In what follows, we motivate this problem setting (Sec.~\ref{sec:motivation}), and formalise it through a dynamic systems perspective (Sec.~\ref{sec:method}) to simulate varying levels of miscommunication (Sec.~\ref{sec:simulations}). Our simulations illustrate that effective communication hinges on the alignment of generative models across different interlocutors. Misunderstandings arise when there is a misalignment between these models, a phenomenon often exacerbated by the linguistic differences (i.e., prior encoding) between native and non-native speakers. By examining the interplay between precise and flexible priors, we gain insight into the multistable dynamics of dyadic communication and the computational mechanisms underpinning effective exchanges. This investigation elucidates the importance of similar generative models for achieving mutual understanding. We conclude with a brief discussion on how flexible priors may alleviate communication misunderstandings by facilitating the convergence towards shared narratives (Sec.~\ref{sec:conclusion}).

\section{Dyadic exchanges and Bayesian inference}\label{sec:motivation}
% We want to look at dyadic exchanges
% based on recent experiment evidence on naturalistic experiments we can see that inter-subject correlations dont happen at the lower sensory regions but are observed across the entire hierarchy of regions implicated in language processing.
Recent experiments have shed light on the nature of dyadic exchanges in language processing~\citep{chen2017shared,baldassano2017discovering,zada2024shared}. Notably, inter-subject correlations are not primarily observed in lower sensory regions but manifest across the entire hierarchy of areas involved in language processing~\citep{stephens2010speaker,dikker2014same,silbert2014coupled}. This finding underscores the complexity of neural synchronisation during linguistic interactions, suggesting that multiple levels of cognitive processes are crucial in dyadic exchanges.

\subsection*{Hierarchical processing of speech}
Understanding the hierarchical processing of speech is crucial to elucidating how the brain integrates multiple levels of linguistic information~\citep{jackendoff2003precis,friederici2002towards,price2010anatomy,sajid2023degeneracy}. The primary auditory cortex is the initial cortical region that receives auditory input and is responsible for processing fundamental acoustic properties such as frequency, intensity, and temporal aspects of sound~\citep{kaas2010evolution}. These serve as the building blocks for more complex auditory perceptions. Information is propagated to higher-order auditory areas, including the superior temporal gyrus, and superior temporal sulcus, which are involved in phonemes recognition~\citep{hickok2007cortical}. The progression from phoneme recognition to word and sentence comprehension involves a network of regions including the inferior frontal gyrus (IFG). The IFG is critical for syntactic processing, enabling the brain to parse and understand the grammatical structure of sentences~\citep{friederici2011brain}. This region is also involved in managing the hierarchical structure of language, facilitating the integration of words into coherent syntactic and semantic frameworks~\citep{hagoort2005broca}. Importantly, the hierarchical processing of speech is not linear but involves intricate feedback mechanisms. Higher-order regions send predictive signals back to lower-order areas to influence the processing of incoming auditory information~\citep{friston2010free,rao1999predictive,sajid2023degeneracy}. 

\subsection*{Predictive coding and dyadic exchanges} 
% Our formulation extends the computational model of the internal state synchronisation during dyadic exchanges, as proposed in \cite{friston2015duet}. 
To formalise this hierarchical processing, we extend the Bayesian model of dyadic exchanges proposed by~\citep{friston2015duet}. This model conceptualises linguistic exchange as an interplay between two agents -- a speaker and a listener -- who share an identical internal model of the world. The model comprises a two-level hierarchy: the first level contending with incoming sensory information and the second level integrating and interpreting these signals within the broader context. 

\begin{figure}[htp!]
    \centering
    \includegraphics[width=1\linewidth]{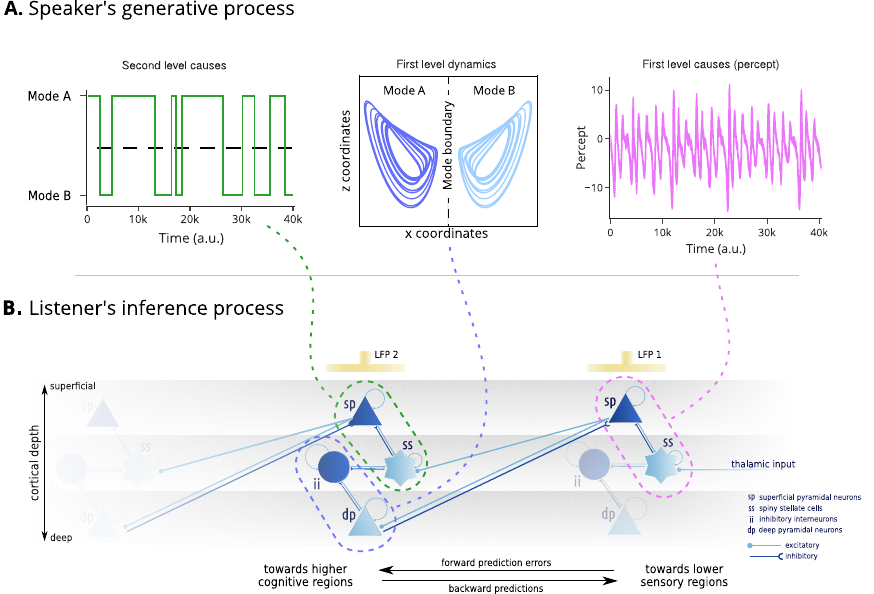}
    \caption{Mechanisms of generation and inference. The upper panel (A) represent the speaker's generative process. This starts from second-level outcomes, determining the mode of the broken Lorenz system. Then, the first level dynamics determines the evolution of the internal states. Finally, the first level outcomes represent the construction of the generated percepts from the internal states. The lower panel (B) denotes the assumed neurobiology of the listener's inference process. The inference process involves two cortical regions whose activity is modelled by a canonical microcircuit capturing the local connectivity between four neuronal populations: superficial pyramidal neurons and inhibitory interneurons (dark blue), encoding prediction errors, and spiny stellate and deep pyramidal neurons (light blue), encoding predictions. The percept (e.g., auditory waveform) is assumed to be conveyed to spiny stellate cells of a lower sensory region through a projection from the thalamus. The discrepancy with the agent's prediction gives rise to prediction errors, encoded through the activity of the superficial pyramidal neurons. Their self-inhibition determines the excitability of the population and is assumed to encode the precision of the predictions. The electrical activity of superficial pyramidal neurons causes local fluctuations in the electromagnetic field potential, assumed to be recorded by an electrode "LFP 1". The superficial pyramidal neurons have excitatory forward connections to higher cognitive regions. Here, the higher cognitive region is responsible for encoding the first-level dynamics of the broken Lorenz system. The prediction errors over the dynamics are encoded by the inhibitory interneurons, whose activity is modulated by the activity of the spiny stellate cells, that encode predictions about the second level outcomes, that is, the mode of the Lorenz system. The prediction errors of the second level model are encoded in the activity of the superficial pyramidal neurons of the second regions, and recorded through "LFP 2". }
    \label{fig:model}
\end{figure}

During an interaction, the speaker's utterances serve as an indicator of their internal beliefs. The listener, equipped with the same model, uses the perceptual cues to align their internal beliefs with the speaker's (i.e., internal state synchronisation) to ensure mutual understanding. This synchronisation process is grounded in hierarchical predictive coding~\cite{friston2008hierarchical}; which posits that perceptual inference occurs through a cascade of predictions and prediction errors across multiple levels of processing. During dyadic exchanges, these hierarchical predictions cover many levels of speech; from low-level acoustic features, sounds, phonemes, and words, to higher-level semantic and syntactic structures. The brain then propagates these predictions backwards (i.e., top-down, from higher cognitive regions to lower sensory regions) and compares them against actual perceptual information. Any differences (i.e., prediction errors) are propagated forward (bottom-up, from lower sensory regions to higher cognitive regions) in the brain after being weighted by the precision (i.e., confidence) of the predictions (Fig.~\ref{fig:model}). This ensures the continual alignment of internal models between the interacting agents, facilitating effective communication.

\textit{Neurobiology} Bayesian inference is realised by updating internal beliefs to explain away the prediction errors. This inference scheme can be related to the neurobiology of the cortex. The activity of superficial pyramidal neurons and inhibitory interneurons correlates with prediction errors, while corresponding predictions are assumed to be encoded by spiny stellate cells and deep pyramidal neurons. The interaction between these neuronal populations forms a canonical microcircuit, responsible for constructing and continuously updating beliefs about the dynamics and states of the world~\citep{bastos2012canonical}. Hierarchical predictive coding emerges from the asymmetry in connections between canonical microcircuits, with superficial pyramidal neurons projecting to higher cortical regions and deep pyramidal neurons sending modulatory projections to lower sensory regions. This can be validated by recording local field potentials (LFP) generated by superficial pyramidal neurons in the cortical or skull surface and comparing them against simulated evoked responses.  

\textit{Empirical evidence} Recent neuroimaging studies have provided evidence for predictive coding in speech perception~\citep{blank2016prediction,sohoglu2012predictive}. For instance, \citep{blank2016prediction} demonstrated that the brain's response to degraded speech is modulated by prior expectations, with increased activity in frontal regions when predictions are violated. Similarly, \citep{sohoglu2012predictive} showed that top-down predictions can override bottom-up sensory input in speech perception, providing evidence for predictive coding. Furthermore, syntactic processing also supports a predictive coding account. Hahne and Friederici~\citep{hahne1999electrophysiological} found that syntactic violations often elicit an early left anterior response, typically occurring between $100-300$ms post-stimulus onset. This is thought to reflect the rapid detection of a syntactic prediction error, although its habituation has been debated~\citep{steinhauer2012early}.

\textit{Generalised filtering in speech processing} While predictive coding provides a powerful framework for understanding speech processing, implementing these ideas requires going a step further. One way to operationalise it is via generalised filtering~\citep{friston2008variational,Friston2010GeneralisedF}, which offers a dynamic Bayesian inference scheme that can be applied to any hierarchical model. For this,  we need to construct a hierarchy of deterministic systems that track the modes of Gaussian distributions that have fixed precision (i.e., inverse variance). The hierarchy can be formalised as a series of interconnected state space models, where each level $i$ is characterised by:
\begin{align} 
        \label{eq:genmodel}
        \frac{dx_i(t)}{dt} &\sim \mathcal{N}\left(f_i(x_i(t), v_{i+1}(t)), \sigma_i\right) \\ v_i(t) &\sim \mathcal{N}\left(g_i(x_i(t), v_{i+1}(t)), \sigma_e\right)~,
\end{align}
where, $x_i$ are states, $v_i$ are the outcomes, $f_i$ the evolution function, $g_i$ an observation function with some additive analytic Gaussian noise at level $i$, $\sigma_i$ the \textit{internal precision} determining the confidence in internal dynamics, and $\sigma_e$ the \textit{external precision} determining the confidence in predicted outcomes. In generalised filtering, the states and outcomes are expressed in generalised coordinates, i.e., by concatenating all derivatives up to a certain order. For further technical details, see~\citep{Friston2010GeneralisedF,friston2008variational}. 

Happily, within the context of dyadic exchanges, generalised filtering can model how listeners continuously update their beliefs about a speaker's internal state based on the ongoing stream of speech input. We rely on this scheme for our simulations (Section~\ref{sec:method}.

\subsection*{Communication between native and non-native speakers} 
Studies have shown that listening to accented speech accentuates activation in brain areas associated with language processing and comprehension. For instance, \cite{adank2012neural} observed enhanced activation in the left inferior frontal gyrus and superior temporal gyrus when native English speakers processed Dutch-accented English, compared to native English speech. This increased neural activity can be interpreted as an indicator of the greater effort required to process and understand accented speech.  From a predictive coding perspective, this increased neural activity likely reflects the brain's efforts to reconcile prediction errors arising from mismatches between expected and actual acoustic patterns in accented speech. Furthermore, \citep{yi2014neural} found that native English speakers listening to Korean-accented English showed increased activation not only in classical language areas~\citep{sajid2022active} but also in regions associated with attention, such as the bilateral insula and right frontal areas. This aligns with the hierarchical nature of predictive coding models. It suggests that when lower-level predictions about phonemes and acoustic features generate an increase in errors, higher-level cognitive processes engage more actively to manage these errors and guide the updating of predictive models across multiple levels of the language processing hierarchy. This cascading effect of prediction errors and model updating may explain the recruitment of additional brain regions beyond the core language network when processing accented speech, reflecting the increased computational demands of adapting to unfamiliar speech patterns.

The neural response to accented speech can differ between native and non-native listeners, potentially influencing communication dynamics. \citep{goslin2012erp} demonstrated that native English speakers exhibited greater activation in the left inferior frontal gyrus when processing foreign-accented speech compared to regional accents, a difference not observed in non-native listeners. From a predictive coding standpoint, this suggests that native speakers' language models generate larger prediction errors when encountering foreign accents, necessitating increased neural activity to update these models. In contrast, non-native listeners' more flexible predictive models, shaped by diverse linguistic experiences, may accommodate accent variations more readily. This difference underscores how the linguistic background of the listener shapes their internal models and plays a crucial role in speech processing.

\section{Methods}\label{sec:method}
To investigate linguistic communication between native and non-native speakers, we explore how multistable dynamics interact with the precision of the internal model during the perception of ambiguous speech segments. For this, we employ a two-stage process to model the speaker-listener interactions: generating the speaker's output signal, and inverting the listener's generative model. 

The \textit{speaker's output signal} is conceptualised as a generative process, governed by a hierarchical probabilistic model. This hierarchical structure reflects the nested nature of language, from phonemes to words to sentences. By integrating a set of differential equations and applying observation functions at each level, we simulate the complex, multi-level process of speech production. This approach captures the idea that speakers generate linguistic outputs based on their internal dynamics, with each level of the hierarchy influencing the levels below.

The \textit{listener's perceptual process} is modelled as inference within a similar hierarchical structure, but crucially, with its own set of priors. We construct this generative model by specifying prior distributions over internal states $x_i$ and outcomes $v_i$ at each level of the hierarchy. The generative model is the joint probability over the data and the internal states. From the listener's perspective, the generative model describes the preferences about its internal configuration --- e.g., particular patterns of neuronal activity --- through priors and preferred states of the sensorium given its internal configuration through the hierarchical model described in \eqref{eq:genmodel}. 

The following sections detail the internal dynamics and the perceptual projections for this hierarchical generative model, the inference scheme and the process for simulating \textit{in-silico} evoked responses.

\subsection*{Bi-stable internal dynamics}\label{sec:methods::internal}
To simulate the internal dynamics we employ a bistable continuous dynamical system: the broken Lorenz system~\citep{li2014multistability}. This model is obtained using a particular set of parameters in the equations of the classical Lorenz system~\citep{lorenz1963deterministic}. The classical Lorenz system was designed to illustrate the unpredictable nature of weather patterns, but its mathematical formulation makes it suitable for exploring more general complex dynamic behaviours \cite{friston2017computational}. Briefly, the Lorenz system is governed by the following set of differential equations~\citep{lorenz1963deterministic}:
\begin{align}
\label{eq:lorenz}
\begin{cases}
\dot{x} &= \sigma (y - x) \\
\dot{y} &= x(r - z) - y \\
\dot{z} &= xy - bz
\end{cases}
\end{align}
where $\sigma$, $r$, and $b$ are parameters that control the system's dynamics. 

The \textit{classical Lorenz system}, for certain parameter values, exhibits chaotic behaviour characterised by its iconic butterfly-shaped attractor. This attractor consists of two lobes, with trajectories swirling around each lobe before intermittently jumping to the other lobe. 

The \emph{broken Lorenz system}, as investigated by~\citep{li2014multistability}, modifies this dynamic by introducing a regime of parameters that disrupts the connection between the two lobes of the attractor. More precisely, by letting $\sigma = 0.12$, $\beta=-0.6$, and $r = 0$ in~Eq.~\eqref{eq:lorenz}, the dynamics remain chaotic but the unique attractor of the original system is divided into two unconnected attracting sets with disjoint basin of attractions. In this altered system, trajectories initialised in one lobe remain confined to that lobe, effectively disconnecting the "wings of the butterfly" (Fig.~\ref{fig:summary}C). This bi-stability creates a scenario where the system's state is constrained to one of two possible attractors, depending on the initial conditions.

While the broken Lorenz system is not \textit{per-se} used by the brain in dyadic exchanges, its  multistable aspect provides an ideal test-bed for investigating mutual understanding in linguistic exchanges. Indeed, each lobe of the bistable system can be seen as distinct internal "speech" models attempting to predict the hidden dynamics of the speech percept. This structural multistability could arise, for instance, from the competition for attention by two or more different neuronal populations within the same region \cite{pastukhov2013multi, gershman2012multistability,sterzer2009neural}. The winning mode then constitutes an intermediate "cause", which would need to be explained away by a higher level region within the cortical hierarchy, evolving over a slower temporal scale~\cite{medrano2024linking}. Under this model, a mutual understanding between the speaker and the listener implies that both use the same internal model to predict the percept, i.e., their internal states evolve on the same lobe of the broken Lorenz system. This hierarchical synchronisation, at the level of both the internal dynamics and the attracting set, mimics the hierarchical aspect of the inter-subject synchrony between speaker and listener observed in naturalistic neuroimaging \cite{silbert2014coupled,thiede2020atypical,li2024speaker}. Similarly, a misunderstanding arises from the listener predicting the speaker's speech while being on the opposite lobe of the Lorenz system. This makes the model ideal for evaluating how the listener's confidence  -- or lack thereof --  over their speech predictions influences their ability to infer the lobe that generated the percept.

\subsection*{Ambiguous outcome}\label{sec:methods::perceptual}
We extend the model for our purposes by introducing an observation function that projects the 3-dimensional state-space trajectories onto a single perceptual outcome:
\begin{align}
v_0 = 0.5 y + z
\end{align}
The projection from internal states to the percept purposely introduces perceptual ambiguity. The percept is computed as the sum of the $z$ coordinates of internal states, over which none of the modes is distinguishable, and a down-weighted $y$ coordinate, over which the modes are partially distinguishable. However, it does not involve the x coordinates, over which the two lobes of the broken Lorenz system are distinguishable (Fig.~\ref{fig:summary}D). Thus,  percepts generated from different lobes of the broken Lorenz system exhibit only sensible differences. Introducing this perceptual ambiguity in the generative model is fundamental to mimic how listeners might struggle to infer the exact meaning of ambiguous spoken words, for instance, the homophones "ship" and "sheep",  when pronounced with an unfamiliar accent.

\subsection*{Relevance for dyadic exchanges}

% I didnt like the previour structure: We need to have introduced hte ambiguous percept to be able to talk about misunderstanding. 

% The broken Lorenz system, under prior beliefs, can lead to divergent but stable states of understanding, akin to being stuck on one side of a conceptual divide (Fig.~\ref{fig:summary}C).
By employing this model, we simulate how different priors (i.e., generative model parameterisation) lead to distinct interpretations (i.e., different attractors) of the same sensory data. This setup allows us to explore how prior beliefs influence the alignment of generative models and the subsequent understanding of percepts. The bi-stable nature of the system serves as an ideal framework to investigate how misunderstandings arise when interlocutors operate under different model assumptions, akin to being on different lobes of the attractor. This is illustrated in Fig.~\ref{fig:summary}, which shows the projection of the Lorenz attractor and highlights the perceptual ambiguity between the speaker and listener over a particular stimulus -- in this example the word "sheep". 

Furthermore, the multistable aspect of the broken Lorenz system provides an ideal test-bed for investigating linguistic exchanges between individuals with different language proficiency. Indeed, different initial conditions mimic prior beliefs and are useful for modelling how different priors can lead to distinct, stable states of understanding. These mirror how individuals can persist in their interpretations of ambiguous linguistic inputs, leading to potential misunderstandings. For instance, two interlocutors with different native languages might interpret the same phrase in distinct ways, resulting in a stable yet divergent understanding based on their linguistic backgrounds and prior experiences. In other words, this model allows us to explore how the listener's confidence (i.e.,  precision) in their language ability influences their capacity to transcend their divergent initial (i.e., prior) beliefs and synchronise to the correct mode of understanding, i.e., the same lobe as the speaker. 

Interestingly, this model represents a hierarchical structure -- akin to those present in linguistic communication~\citep{friston2021active,sajid2022mixed} --with a second-level semantic encoding (i.e., lobe) and a first-level continuous encoding (i.e., the state within a lobe). This hierarchical representation is crucial for capturing the layered nature of linguistic processing, where high-level semantic meanings (corresponding to which lobe the system is in) guide the interpretation of continuous, lower-level linguistic features (corresponding to the specific state within a lobe). This structure allows us to explore how overarching semantic frameworks influence the detailed parsing and interpretation of linguistic inputs.

\begin{figure}[!t]
    \centering
    \includegraphics[width=\textwidth]{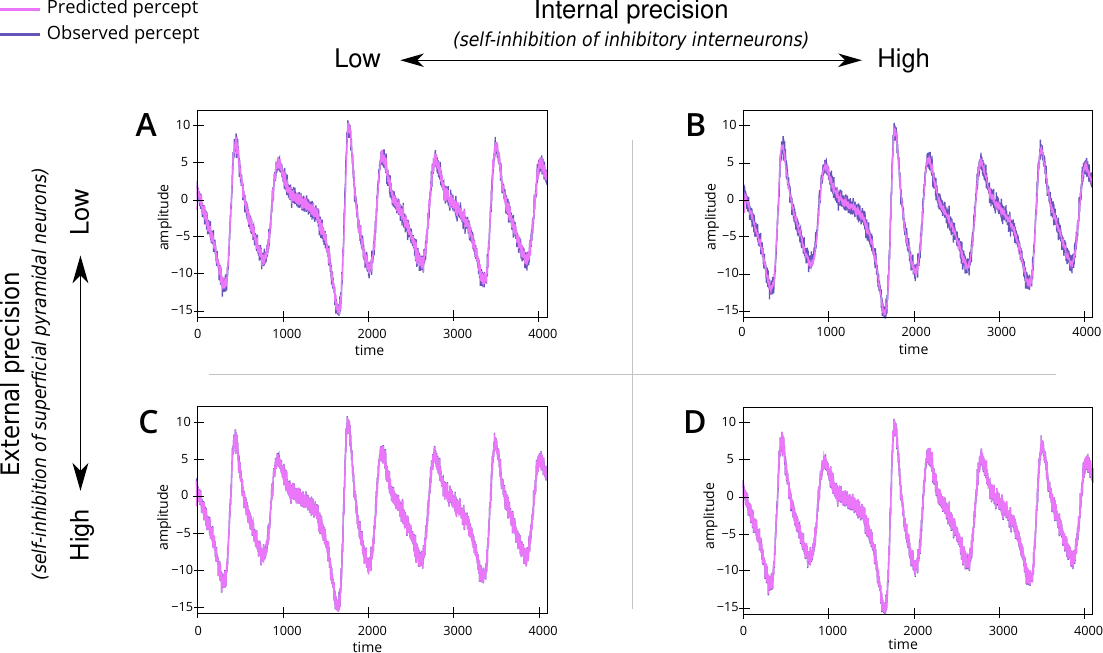}
    \caption{Line plots (A-D) show the percept across time for the 4 simulations. A is the baseline simulation, B is the simulation with high internal precision and moderate external precision, C has moderate internal precision and high external precision and D has high internal and external precision. For each plot, the y-axis reveals the amplitude and the x-axis is the time measured on some arbitrary scale. Here, the "observed percept" (in purple) is generated by the speaker and the "predicted percept" (in pink) denotes the inference made by the listener.}
    \label{fig:percept}
\end{figure}

This formulation provides a nice opportunity to study the effect of multi-stability on hierarchical variational inference. In particular, it provides a computational account of how different inferences, modelled here as convergence towards different lobes of the attractor, can arise given the same percept (Fig.~\ref{fig:percept}\footnote{Note, the blue line (stimulus) is defined as $v_0 = 0.5y_{\text{s}} + z_{\text{s}}$, where $y_{\text{s}}$ and $z_{\text{s}}$ are coordinates in the speaker's broken Lorenz system. The red line (percept) is the estimated $\Tilde{v}_0$; inferred via generalised filtering~\citep{Friston2010GeneralisedF}.}).

\subsection*{Inversion scheme}
Having specified the generative model, one could apply Bayes rule to obtain a \textit{posterior distribution} of the internal states and outcomes, i.e., their distribution refined from the sensory data. However, exact Bayes inference is computationally -- or neuroanatomically -- intractable and a solution must be approximated. Thus, we rely on generalised filtering -- a Bayesian filtering algorithm that enables continuous data assimilation~\citep{Friston2010GeneralisedF} (Sec.~\ref{sec:motivation}) -- to estimate the posteriors. 

This is done by constructing a hierarchical dynamical system, with the same number of levels as Eq.~\eqref{eq:genmodel}, and evolving against the gradient of the generative model's free energy $F$. This free energy, known as evidence lower bound (ELBO) in machine learning, scales with the distance between the approximate and exact solution to the Bayesian inference problem. The free energy gradients correspond, under the Gaussian model, to precision-weighted prediction error. In other words, the system flows against prediction errors with a rate determined by precision. Here, the free energy minima gives an approximate solution to the Bayesian inference problem. The solution is hierarchically structured, involving the construction of top-down predictions, comparison of lower-level predictions with sensory observations, and upward propagation of resulting prediction errors through the hierarchy. This update process allows for identifying the most likely cause of perceptual data under the current beliefs by solving an approximate Bayesian inference problem. Interestingly, this mathematical model yields an algorithmic construct that closely mimics aspects of hierarchical predictive coding in the brain \cite{friston2008hierarchical,friston2010free}. 

\subsection*{Modelling in-silico evoked responses} 
We can use the connection between generalised filtering and hierarchical predictive coding to generate \textit{in-silico} LFPs by looking at the precision-weighted prediction errors at each level of the hierarchical model as the average activity of a population of superficial pyramidal neurons in a canonical microcircuit model, following~\citep{auksztulewicz2016repetition} (Fig.~\ref{fig:model}). This allows us to probe the mechanisms underlying linguistic misunderstandings and their neural correlates.

For this, we augmented the model with a lobe variable acting as a second-level "cause" (see Fig.~\ref{fig:model}). During the generation of the percept, this variable $v_1$ is switched between 1 and -1 and used to give the sign of the y-component in the percept equation: 
\begin{align}
        v_0 = 0.5 \, v_1 \, y + z \text{ with } u \in \{-1,1\}.
\end{align} 

This exploits the symmetry of the Lorenz system over the $x-y$ axis to simulate an instantaneous lobe switching. 
    
To allow the listener to infer the lobe used by the speaker, we integrate another level that tracks the lobe variable, i.e., we let the listener hold beliefs about the current lobe. These beliefs are static (i.e., the predicted dynamics of the lobe variable is 0), and used in the first-level generative model as: 
\begin{align}
        \Tilde{v}_0 = 0.5 \, \tanh(\Tilde{v}_1) \,\Tilde{y}  + \Tilde{z}
\end{align}

where $\tanh$ is the hyperbolic tangent function that maps a real variable between $-1$ and $1$ and the tilde notation is used to denote the listener's variables. The listener's priors on the lobe variable are Gaussian with mean $1$ and log-precision $-1$. This intermediate precision allows us to model a listener who is confident that Mode A (Fig.~\ref{fig:summary}) was used to generate the data but is receptive to gradually revising this belief based on accumulating evidence.

Our ERP simulation of the listener's inference process employs the following procedure. First, we generate a percept of 16,384 points using the speaker's generative process. Next, we invert the listener's model using DEM, producing a prediction of the percept, internal states (i.e., the location within a lobe of the broken Lorenz system) and lobe variable for each time point. The prediction error, corresponding to the difference between prior predictions and posterior estimates, is then multiplied by the precision of the internal beliefs. The precision-weighted prediction errors over the percept are taken as LFPs for the "first region". This reflects the neurobiological assumption of canonical microcircuit models that superficial pyramidal neurons, who contribute predominantly to LFPs, encode the precision-weighted prediction errors of the outcomes. Similarly, the LFP of the 'second region' is approximated by the precision-weighted prediction errors associated with the lobe variable. The LFPs are then extracted on a peri-stimulus window of -5 points to 25 points around a lobe switch, up-sampled 10 times using cubic interpolation\footnote{We assumed that introduced high-frequency artefacts are attenuated by averaging as not time-locked to the stimulus onset, while low-frequency components are preserved.}, separated between transitions from the unexpected lobe to the expected one (from -1 to 1) from transitions to the unexpected lobe (from 1 to -1), and finally averaged over stimulus repetition.

\section{Results}\label{sec:simulations}
\begin{figure*}[!t]
    \centering
    \includegraphics[width=\textwidth]{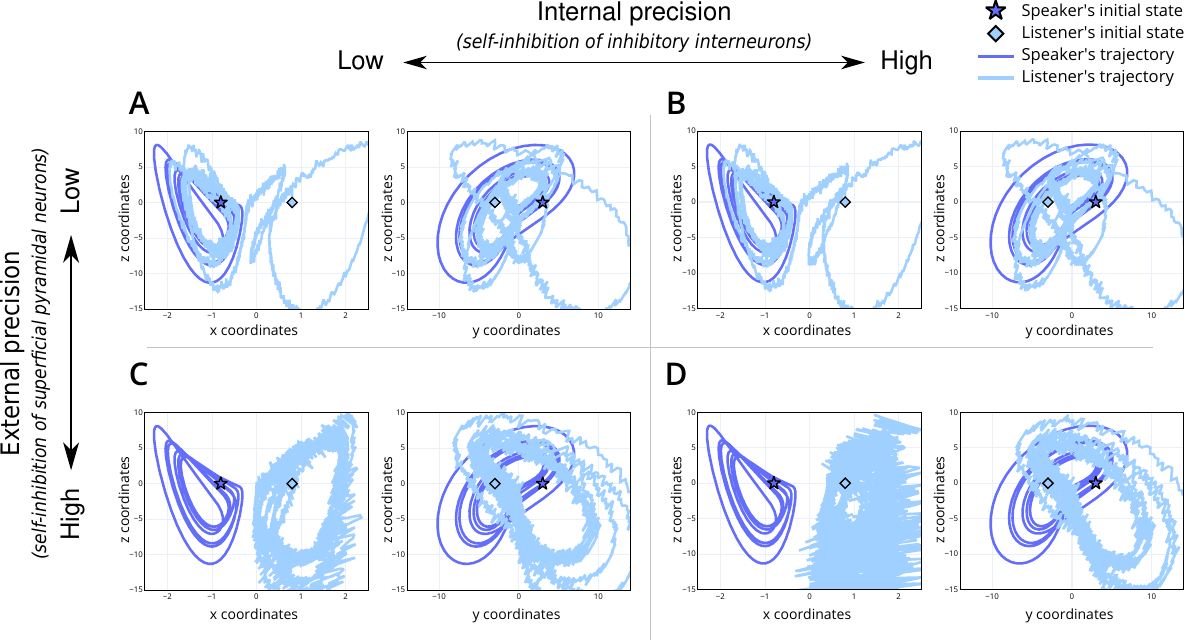}
        \caption{Trajectory of the speaker and listener's internal dynamics for the 4 simulations. A is the baseline simulation with moderate internal and external precision, B is the simulation with high internal precision and moderate external precision, C has moderate internal precision and high external precision and D has high internal and external precision. For each plot, the y-axis is the z-coordinate of the Lorenz system and the x-axis is the x-coordinate (left within each panel) or y-coordinate (right within each panel). Here, the dark blue line denotes the speaker's trajectory and the light line denotes the listener's. The initial state of the speaker and listener are denoted by a dark blue star and a light blue diamond, respectively.  }
    \label{fig:xz}
\end{figure*}
To simulate linguistic exchanges between native and non-native individuals, we use the broken Lorenz system introduced in the previous section (Sec.~\ref{sec:method}). We manipulate the precision over the prior internal states and external causes (Tab.~\ref{tab:precision}). This allows us to explore how variations in prior beliefs influence the alignment of generative models and the resulting states of potential misunderstanding. Our simulations are designed to represent different interaction scenarios by varying the precision of priors in both internal dynamics (analogous to linguistic structure and semantics) and external causes (analogous to sensory inputs and context). By systematically adjusting these parameters, we can simulate the responses of native and non-native listeners to non-native speech under various communicative conditions. For each simulation, the model is initialised with the same conditions: $(0.8, -3, 0)$ for the speaker and $(-0.8, 3, 0)$ for the listener. Data is then inverted using generalised filtering~\citep{Friston2010GeneralisedF}. 

\begin{table}[!ht]
    \centering
    \begin{tabular}{c| c | c }
        Simulation & Internal precision & External precision  \\
        \hline
        \hline
        A & $e^0$ &  $e^0$ \\
        B & $e^4$ & $e^0$ \\ 
        C & $e^0$ & $e^4$ \\ 
        D & $e^4$ & $e^4$ \\ 
    \end{tabular}
    \caption{Overview of the prior internal and external precision}
    \label{tab:precision}
\end{table}
% \begin{figure}[!t]
%     \centering
%     \includegraphics[width=\textwidth]{figures/figure_2.pdf}
%     \caption{Line plots of the internal states (x,y and z) across time for the 4 simulations. A is the baseline simulation, B is the simulation with high internal precision and low external precision, C has low internal precision and high external precision and D has high internal and external precision. For each plot, the y-axis reveals the internal estimate and the x-axis is the time measured on some arbitrary scale. Here, the dashed line denotes the internal beliefs of the speaker and the solid line denotes the internal beliefs of the listener. There are 3 internal states: x in blue, y in orange and z in red.}
%     \label{fig:internal}
% \end{figure}

\subsection*{Low internal and external precision}
We introduce a baseline simulation (Tab.~\ref{tab:precision}; simulation A). For this scenario, we set the internal and external precision to $e^0$. This parameterisation represents a non-native listener, with low confidence in their perceptual abilities and the dynamics of the shared narrative. The simulation results show alignment between speaker and listener (Fig.~\ref{fig:percept}A) and convergence to the appropriate trajectories within the attractor (Fig.~\ref{fig:xz}A) i.e., the correct lobe. This serves as a reference point for understanding the dynamics in subsequent simulations.

% \begin{figure}[!t]
%     \centering
%     \includegraphics[width=\textwidth]{figures/figure_4.pdf}
%         \caption{Plots of the dynamics of the y-z coordinates of the Lorenz system across time for the 4 simulations. A is the baseline simulation, B is the simulation with high internal precision and low external precision, C has low internal precision and high external precision and D has high internal and external precision. For each plot, the y-axis is the y-coordinate and the x-axis is the y-coordinate. Here, the blue line denotes the speaker (i.e., trace 0) and the red line denotes the listener (i.e., trace 1).}
%     \label{fig:yz}
% \end{figure}

\subsection*{High internal precision with low external precision}
%% #1: high internal precision, low external: This is a non-native speaker listening to the non-native speaker. They understand what the speaker means, without too much focusing on the actual phrasing or pronunciation.  
We simulate a situation where a non-native listener, characterised by low confidence in its perceptual ability, has high confidence in the dynamics of the shared narrative, reflected by precise (or confident) linguistic beliefs (e.g., semantics) (Tab.~\ref{tab:precision}; simulation B). This reflects a context where the interacting individuals understand the underlying meaning or narrative but express it differently due to linguistic variations i.e., maybe different ways of pronouncing a particular word. To represent this, we set high internal precision ($e^4$) and the external precision to a low level ($e^0$). 

The simulation results reveal that, despite starting from different stable points within the attractor, the shared semantic understanding allows the system to converge towards a common attractor set (Fig.~\ref{fig:xz}B). This indicates that a shared narrative can bridge linguistic differences, facilitating mutual understanding even when vocabulary or expression varies significantly. The attractor convergence is visually represented in the trajectories within the Lorenz attractor (Fig.~\ref{fig:xz}B). Heuristically, this can be thought of as understanding arising from finishing the other person's sentences by continuing the interaction i.e., a common belief space with different outcomes or vocabulary. In terms of physiology, we would anticipate an attenuation of the prediction error as the linguistic exchange continues. This reduction reflects the system's ability to reconcile perceptual discrepancies through a shared narrative. The dynamics observed here highlight a key aspect of bilingual communication: the ability to maintain a coherent exchange through shared semantic grounding, even when surface linguistic features differ. 

\subsection*{Low internal precision with high external precision}
%% #2: low internal precision, high external: Overly precise over the percept, not so sure about the meaning. That represents a native speaker hearing to a nonnative speaker. Maybe low internal precision results from accumulated gramatical errors. The native speaker is thus paying very strong attention to the percept, but that does not help them find the right lobe.  
We simulate a situation where a native listener, characterised by precise (or confident) percepts (e.g., pronunciation), interacts with a non-native speaker (Tab.~\ref{tab:precision}; simulation C). This reflects a context where the native listener tries to infer the underlying narrative based on the linguistic input. To simulate this, we set the internal precision to $e^0$ (same as baseline) while increasing the external precision to $e^4$ i.e., overly confident priors of how words should be articulated. 

The simulation results illustrate that the system can interpret the percept (Fig.~\ref{fig:percept}C) but struggles to find the correct attractor (Fig.~\ref{fig:xz}C). That is, there is divergence in the semantic understanding despite linguistic synchronisation. This misalignment is evident in the trajectory plots (Fig.~\ref{fig:xz}C), where the system fails to localise within the correct attractor basin, illustrating the native speaker's difficulty in adjusting to unexpected linguistic inputs from the non-native speaker and inability to resolve the uncertainty over the true attractor set. (Fig.~\ref{fig:xz}C) demonstrate that the trajectories, instead of converging to a shared attractor, remain dispersed. This dispersion indicates the system's inflexibility and the resultant divergence in understanding despite continued interaction -- perhaps, due to increased accumulated grammatical errors. Regarding physiology, we would anticipate a prolonged increase in the prediction error as the linguistic exchange continues. This persistent prediction error should underscore the native speaker's ongoing struggle to integrate unexpected linguistic features with their internal model. The high external precision effectively localises the system into a rigid inference regime (i.e., the incorrect attractor set), unable to adapt to the non-native speaker's variable inputs.

\subsection*{High internal and external precision}
% Simulation 3.  (outer-inner obtuse model)
We simulate a scenario where the naive listener is confident about the  dynamics of the shared narrative  (Tab.~\ref{tab:precision}; simulation D). For this, we parameterise internal and external precision to high levels, i.e., $e^4$. 

The model fails to converge to the right lobe (Fig.~\ref{fig:xz}D) and exhibits inaccurate internal beliefs (Fig.~\ref{fig:xz}D), despite its capacity to accurately infer the percept (Fig.~\ref{fig:percept}D). Importantly, the trajectory remains erratic and dispersed, failing to stabilise within the attractor basin (Fig.~\ref{fig:xz}D). This dispersion indicates that the high precision in both internal and external states leads to over-fitting and excessive sensitivity to perceptual inputs, preventing the system from achieving stable, coherent interpretations. Therefore, an overabundance of precision can be detrimental to effective inference about the causes of the states of the world. Heuristically, this scenario could represent individuals who are overly fixated on precise linguistic rules and sensory details, leading to constant re-evaluation and reinterpretation of inputs. Such rigidity and sensitivity hinder the establishment of a stable shared narrative, resulting in persistent misunderstandings and communication breakdowns. Physiologically, we would anticipate a persistent and elevated prediction error throughout the linguistic exchange, reflecting the system's ongoing struggle to stabilise its internal model. This persistent prediction error underscores the interlocutors' difficulty in integrating high-precision expectations with the variable linguistic inputs. The high internal and external precision effectively localises the system into a volatile inference regime, unable to adapt to the inherent variability in natural language exchanges.

\begin{figure}[ht!]
    \centering
    \includegraphics[width=0.5\textwidth]{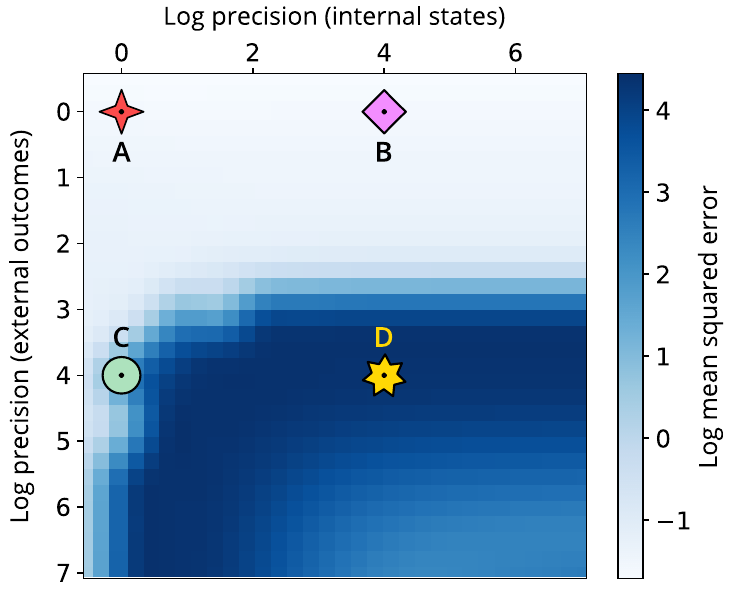}
    \caption{Heatmap showing the log mean squared error (MSE) across varying internal and external precision (measured in log units). The MSE is over the internal states: x, y and z. Here, low values denote a synchronisation to the correct lobe, high values denote convergence towards the incorrect (initial) lobe. The four different symbols indicate the four simulations: low internal and external precision (red 4-branch star), high internal and low external precision (pink diamond), low internal and high external precision (green circle), and high internal and external precisions (yellow star).  }
    \label{fig:precision}
\end{figure}

\subsection*{Trade-off between internal and external precision}
To provide a complete picture, we examined how varying internal and external precision influences the effectiveness of perceptual inference. Fig.~\ref{fig:precision} presents the mean squared error (MSE) in log scale across different precision parameterisations for the internal state space. Our analysis highlights that shifts in internal precision alone have little impact on the quality of perceptual inference. Conversely, increasing external precision reveals a significant rise in MSE, indicating greater volatility and instability in the system’s internal beliefs. However, this declines for certain regimes where internal precision is also higher. 

% Two scenarios:: 
% - overly precise priors can hinder your optimisation 
% - tighter bound on what is acceptable - then can't converge
% - wider precision - then can converge

% -- both can explain percept but one has meaningless state space

% non native speakers have larger speakers
% precision influences the schronisations

\subsection*{Simulated in-silico evoked responses} 
    
\begin{figure}[t!]
    \centering
    \includegraphics[width=1\linewidth]{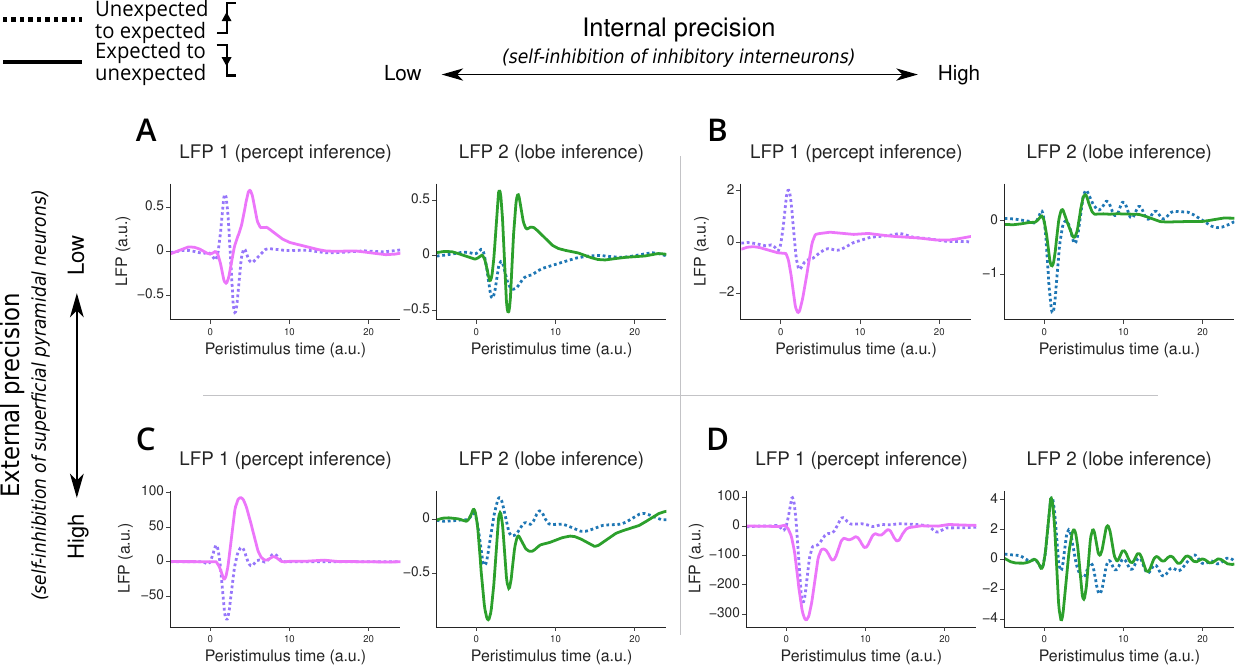}
    \caption{Error-related potential simulated under the two-region model. Panel A corresponds to the model with low internal and external precision, panel B to the model with high internal and low external precision, panel C to the model with low internal and high external precision and panel D to the model with high internal and external precision. Each plot displays an error-related potential as a function of peristimulus time when transitioning from the unexpected to the expected lobe (dotted line) or from the expected to the unexpected lobe (plain line). The LFP value, on the y-axis, corresponds to the precision-weighted prediction error of the causes in the first region (responsible for percept inference, on the left plot in each panel) and the second region (responsible for lobe inference, on the right plot in each panel). The peristimulus time corresponds to time steps after the lobe switching. The displayed ERP is upsampled 10 times using cubic interpolation.}
    \label{fig:erps}
\end{figure}

Our simulations reveal interesting patterns in the ERPs during the listener's inference process after switching between different "lobes" or modes (Fig.~\ref{fig:erps}). First, we observe that when transitioning from unexpected to expected speech patterns, the listener's perceptual system adapts more quickly (i.e., return to 0) than when transitioning to unexpected patterns. This rapid adaptation doesn't necessarily mean the listener has identified the correct mode, but rather that they've managed to predict the percept.

The model's behaviour changes significantly with different parameter settings. When we increase the external precision -- which corresponds to reducing self-inhibition in superficial pyramidal neurons -- we see a marked increase in the amplitude of error-related potentials in the first processing region. This effect is less pronounced in the second region, highlighting the specialised roles of different hierarchical levels. This high external precision scenario mimics a native listener's neural response, where violations of accent or grammar typically elicit stronger ERPs compared to non-native listeners.

Interestingly, we notice that changes to the internal precision have a relatively small influence on the ERP amplitudes compared to the external precision perturbations. The difference in impact between external and internal precision arises from their distinct neural targets. External precision directly modulates the activity of superficial pyramidal neurons, which are the primary contributors to our measure of LFPs. In contrast, internal precision primarily affects inhibitory interneurons, which have minimal direct influence on LFPs. Thus, the effect on the LFP is indirect and mediated through changes in the inference processes. This is reflected through changes in the amplitudes of the ERPs in both regions, as well as a strong change in the overall shape of the ERPs.

\section{Discussion}\label{sec:conclusion}
%  Overview:
Here, we explored the computational mechanisms underlying linguistic miscommunication between native and non-native speakers using the broken Lorenz system. By manipulating internal and external precision, we simulated various scenarios reflecting different levels of linguistic proficiency. We provide insights into how generative model parameterisation may influence communication dynamics and underscore the importance of appropriate belief representations in achieving mutual understanding, i.e., convergence towards a shared narrative.

% Summary of results:
Our simulations reveal that the precision of internal and external priors significantly impacts the alignment of generative models and the resultant states of understanding. Native speakers, with highly precise external priors, exhibit rigid expectations about linguistic inputs. This rigidity leads to difficulties in adjusting to unexpected variations, such as unconventional syntax or pronunciation errors introduced by non-native speakers. Consequently, this misalignment manifests as persistent prediction errors and communication breakdowns. In contrast, non-native speakers, characterised by precise internal priors, demonstrate greater flexibility in adapting to novel inputs. This adaptability facilitates better alignment of generative models, promoting more effective communication despite linguistic differences.

The results from simulations with high precision in both internal and external states highlight the detrimental effects of excessively high precision. This is because high precision may lead to instability and over-fitting; the system becomes overly sensitive to minor variations in perceptual inputs, preventing stable convergence and coherent understanding. This scenario underscores the importance of meta-control over precision priors to maintain stable communication dynamics.

% ERPs
The differential effects of external and internal precision on ERPs offer a potential mechanistic explanation for the observed differences between native and non-native listeners in processing accented speech~\citep{goslin2012erp,yi2014neural}. Native listeners, represented by high external precision in our model, show stronger error signals when encountering unexpected inputs (e.g., foreign accents). This could be interpreted as their highly tuned models generating larger prediction errors that require more neural resources to resolve~\citep{sohoglu2012predictive}. Conversely, the more muted response in the model with lower external precision aligns with empirical observations that non-native listeners often show reduced sensitivity to accent-related variations~\citep{goslin2012erp,bradlow2008perceptual}.

Our model can account for well-known ERP components associated with language processing, specifically the N400 and P600. The N400, typically linked to semantic processing, can be interpreted in our model as reflecting semantic-level prediction errors. Our framework predicts larger N400 responses in native speakers when encountering semantically incongruous words compared to non-native speakers, consistent with empirical findings~\citep{weber1996maturational,hahne2001s}. Similarly, the P600, associated with syntactic processing, can be viewed as reflecting syntactic prediction errors and reanalysis. Our model predicts larger P600 responses in native speakers for syntactic violations. This is consistent with studies showing larger P600 effects in native speakers compared to non-native speakers ~\citep{hahne2001processing}. The lower external precision in non-native listeners' models would result in smaller prediction errors overall, potentially explaining the frequently observed attenuation of both N400 and P600 effects in this population~\citep{ojima2005erp,steinhauer2009temporal}.

% Hierarchy: 
Our simulations highlight the hierarchical nature of linguistic processing, where high-level semantics guide the interpretation of lower-level features. This hierarchical structure is fundamental to understanding how complex language comprehension and production interplay~\citep{jackendoff2003precis}. At the top of this hierarchy, high-level semantics encompass broad, abstract concepts and contextual information, forming the cognitive backbone for interpreting incoming linguistic data~\citep{friederici2002towards,pulvermuller2002neuroscience}. These act as templates that shape our expectations and interpretations, ensuring that we can make sense of diverse and often ambiguous linguistic inputs. As these high-level features influence the processing of lower-level features, such as syntax, phonetics, and individual word meanings, they enable the dynamic integration of these elements into coherent, meaningful communication~\citep{friston2021active,friston2020generative}. This top-down influence ensures that even when specific linguistic inputs are unclear or ambiguous, our overall understanding remains intact by relying on the broader context provided by these high-level structures.

This highlights that effective communication is not solely about the accurate inference about words or sounds but also involves the high-level alignment of underlying narratives between interlocutors. By adjusting their generative models in response to new and variable inputs, speakers can better align their internal representations, facilitating smoother and more accurate exchanges of meaning. For example, non-native speakers may rely heavily on high-level semantic frameworks to compensate for gaps in their understanding of lower-level linguistic features. Conversely, native speakers with more precise models must learn to adjust their expectations and interpretations to accommodate the linguistic variations introduced by non-native speakers.

% Learning
This suggests that adjusting priors in response to new and variable inputs is important for achieving mutual understanding~\citep{chomsky2005three,jackendoff2003precis}. This adaptability is particularly important in multilingual contexts, where linguistic inputs are inherently diverse and unpredictable. The results emphasise that mutual understanding is facilitated by generative models that strike a balance between being sufficiently confident (precise enough) but flexible, allowing interlocutors to navigate the complexities of a particular interaction. One potential approach to enhancing this adaptability is the introduction of learning mechanisms, enabling individuals not only to infer the linguistic percept given their generative model but also to update the model in alignment with incoming inputs. For instance, native speakers exposed to a new linguistic environment, such as a native English speaker moving to France, may fine-tune their external precision to accommodate different stimuli more effectively.

% Learning regime differences: 
Practically, this would entail equipping our current model with a learning scheme -- moving beyond fixed internal and external precision. Here, we consider two different types of learning regimes that could be implemented. Under the first regime, both the speaker and listener would continuously update their precision till synchronisation of their internal beliefs. This can be seen as slowly exercising a change in the dynamical stability of the coupled system and would entail a separation of temporal scales and an abrupt switch to synchronisation~\citep{haken1985theoretical,haken2021information,medrano2024linking}. The second regime would consider a slow learning procedure where updates post a particular interaction -- akin to the meta-learning schemes considered in the literature~\citep{vilalta2002perspective,pmlr-v70-finn17a}. Future research should incorporate a more sophisticated inference procedure that updates the precision over priors across time.

These precision-weighted dynamics exemplify a fundamental principle that extends across multiple domains. This balance between precise and imprecise inference reflects the brain's inherent need to optimise its internal model while maintaining adaptability to novel inputs. This dichotomy can be understood as a trade-off between minimising prediction errors and maintaining model flexibility. Native speakers exhibit higher precision, reflecting years of experience-dependent learning. This results in more efficient processing but potentially reduced flexibility to unfamiliar inputs. Conversely, non-native speakers operate with lower precision, allowing for greater plasticity in their generative models but at the cost of increased uncertainty. This precision balancing act is ubiquitous in learning. During critical periods, sensory systems maintain low precision to facilitate rapid learning, gradually increasing precision as optimal models are established. Similarly, skill acquisition in any domain involves a progression from imprecise, attention-demanding processes to precise, automated routines~\citep{fountas2020deep,yuan2023hierarchical}.

% Limitations:
The primary limitation of this work lies in the idealised setting provided by the broken Lorenz system. This model simplifies the complexities of real-world linguistic interactions, which involve numerous factors such as emotion and social cues, not captured by our model~\citep{tanenhaus2006eye}. Despite this, our study provides a step toward understanding the computational mechanisms of linguistic miscommunication through the lens of dynamic systems theory. Addressing the identified limitations in future research will be essential for understanding linguistic miscommunication. By incorporating more complex features (linguistic or otherwise), and validating with empirical data, future work can build on our findings to create a more comprehensive model of multilingual communication.

\section*{Code availability} The results can be reproduced using the simulations provided here: \href{https://github.com/johmedr/dempy}{https://github.com/johmedr/dempy
}

\section*{Supporting information}{The Wellcome Centre for Human Neuroimaging is supported by core funding from Wellcome [203147/Z/16/Z]. NS was funded by the Max Planck Society.}

\newpage
\bibliographystyle{apalike} 
\bibliography{references}

\end{document}